# Real-time volumetric free-hand ultrasound imaging for large-sized organs: A study of imaging the whole spine

Caozhe Li[1,#], Enxiang Shen[1,#], Haoyang Wang[1], Yuxin Wang[1], Jie Yuan[1,*], Li Gong[2], Di Zhao[2], Weijing Zhang[2], Zhibin Jin[2]

1. School of Electronic Science and Engineering, Nanjing University, China (Corresponding author: Jie Yuan, e-mail: yuanjie@nju.edu.cn).
2. Affiliated Drum Tower Hospital, Medical School of Nanjing University, China.

\* Corresponding author: Jie Yuan
\# Caozhe Li and Enxiang Shen contribute equally in this study.

*Abstract*—Three-dimensional (3D) ultrasound imaging can overcome the limitations of conventional two-dimensional (2D) ultrasound imaging in structural observation and measurement. However, conducting volumetric ultrasound imaging for large-sized organs still faces difficulties including long acquisition time, inevitable patient movement, and 3D feature recognition. In this study, we proposed a real-time volumetric free-hand ultrasound imaging system optimized for the above issues and applied it to the clinical diagnosis of scoliosis. This study employed an incremental imaging method coupled with algorithmic acceleration to enable real-time processing and visualization of the large amounts of data generated when scanning large-sized organs. Furthermore, to deal with the difficulty of image feature recognition, we proposed two tissue segmentation algorithms to reconstruct and visualize the spinal anatomy in 3D space by approximating the depth at which the bone structures are located and segmenting the ultrasound images at different depths. We validated the adaptability of our system by deploying it to multiple models of ultra-sound equipment and conducting experiments using different types of ultrasound probes. We also conducted experiments on 6 scoliosis patients and 10 normal volunteers to evaluate the performance of our proposed method. Ultrasound imaging of a volunteer spine from shoulder to crotch (more than 500 mm) was performed in 2 minutes, and the 3D imaging results displayed in real-time were compared with the corresponding X-ray images with a correlation coefficient of 0.96 in spinal curvature. Our proposed volumetric ultrasound imaging system might hold the potential to be clinically applied to other large-sized organs.

## I. Introduction

ULTRASOUND is extensively utilized in medical imaging due to its notable advantages of being rapid, convenient, and cost-effective. Compared to the conventional B-mode ultrasound, three-dimensional (3D) ultrasound has the capacity to overcome limitations in structural observation and measurement [1]. By merging multiple cross-sectional images, it enables comprehensive and detailed visualization of anatomical structures, leading to enhanced accuracy and precision in diagnosing various medical conditions. Therefore, researchers and medical practitioners have been exploring and implementing volumetric ultrasound imaging in clinical areas such as the carotid arteries [2-4], thyroid [5,6], kidney [7,8], heart [9-11], and breast [12].

However, for large-sized organs such as the liver, muscle, and spine [13], real-time 3D ultrasound imaging is rather challenging. One of the primary difficulties is the increased time and data volume that needs to be processed in real-time due to the organ's size. This requires not only high computational performance but also poses challenges for patients who must maintain immobility for longer periods during the imaging process. Furthermore, the presence of abundant soft tissue surrounding the organ can introduce interference in the identification of organ features, ultimately impacting the physician's observation of the 3D image [14]. To address the above challenges, we designed a real-time volumetric ultrasound imaging system and tested its performance in applying it to the clinical diagnosis of scoliosis.

Scoliosis is a medical condition characterized by a 3D spine deformity, involving lateral deviation and axial rotation of the vertebral column [15]. Adolescent idiopathic scoliosis (AIS) is the most prevalent form of scoliosis affecting about 5% of kids in China [16]. Regular diagnostic imaging, specifically radiography (commonly known as X-ray photography) [17], is essential for both clinical diagnosis and monitoring of patients with AIS [18]. However, Simony et al. [19] discovered that the AIS patients they examined had an overall cancer rate of 4.3%, which was five times higher than the age-matched Danish population. Considering the patient's health, doctors today are consciously reducing the frequency of X-ray usage. Nevertheless, whether undergoing treatment or annual physical examination, adolescents with scoliosis still use X-ray examinations at least every six months until skeletal maturity [18]. In contrast to X-ray photography, 3D ultrasound imaging offers advantages for scoliosis examinations. Firstly, it is a radiation-free imaging technique, eliminating the potential risks of X-ray radiation exposure. Additionally, 3D ultrasound imaging can also provide a 3D view of the spine, enabling better visualization of spinal structures. This enhanced visualization facilitates accurate measurements of the ultrasound curve angle (UCA), a crucial parameter in assessing scoliosis [20,21].

These advantages above motivated researchers to develop systems for 3D ultrasound imaging of the spine. Purnama et al. [22] employed a segmented scanning and stitching method to address the challenges posed by a large amount of data and long acquisition time. However, this approach compromised real-time imaging capabilities and their final 3D imaging lacked an effective way to visualize spinal anatomy. Jiang et al. [23] presented a new method for real-time imaging of the spine. In this study, a rendering method was utilized to directly generate coronal images from the raw image dataset, eliminating the need for reconstructing the 3D volume and resulting in significant time savings. Additionally, the image information was captured from a fixed depth within the skin to obtain spinal anatomy. Victorova et al. [24] used robotic manipulators to automate the 3D ultrasound assessment for scoliosis. However, this system requires the operator to first manually define the path for the robot to follow. There are also many studies based on the Scolioscan (TMIL, Shenzhen, China) [25-27]. But it is important to note that the system uses a custom-designed linear probe with a frequency of 4–10MHz and a width of 10cm [21]. While the wide probe allows for data acquisition to be accomplished with a single-column scan, we found that the large probe struggled to conform well to the patient's uneven back when collecting patient data. Moreover, for some obese people, ultrasound probes with high center frequency have poor imaging quality in deep tissues, which may affect the imaging of spinal anatomy.

Currently, there are two primary approaches for presenting structural information of the spine. The first method involves performing bone surface feature extraction either manually or using trained neural networks after the data acquisition is completed [28-30]. However, manual operation is time-consuming, and neural networks may be influenced by the presence of soft tissues surrounding the bone structure, resulting in unsatisfactory spinal anatomy results. The second method is based on a simple projection method to obtain an anatomical map of the spine by extracting high-brightness or fixed-depth information [31,32]. This method is simple to implement and can therefore be applied to the real-time visualization of spinal anatomy. However, it ignores the interference of some tissue above the bones and the fact that the thoracic and lumbar vertebrae are at different depths from the skin surface.

In this study, we proposed a new approach to deal with the above challenges. Unlike the customized large line array probes utilized in the Scoliscan system, our system uses different ultrasound probes by fixing generic molds. The ability to use various types of probes ensures that the system has the required depth of data acquisition and skin fit. However, the use of small ultrasound probes results in increased imaging time for the same region. To compensate for this issue, we introduced a novel solution for body immobilization that provides more directional restrictions than other studies did [23-27]. We also introduced a real-time algorithm for optical marker recognition, leading to faster localization and a higher sampling frame rate which enables a quicker scanning and reduces data acquisition time. To address the difficulty of processing a substantial volume of data in real-time, we implemented an incremental imaging method coupled with algorithmic acceleration to ensure that the acquired image data can be reconstructed and visualized in real-time. Additionally, to acquire spinal anatomy for clinical diagnosis, intercepting the information at a fixed depth limits the probe to remain in the same direction during the scanning process. We proposed a feature extraction method for dynamically recognizing the depth of the cut that allows free movement during the scanning process. This not only simplifies the requirements for the physician but also optimizes the issue of inconsistency in the depth of the thoracic and lumbar spine relative to the skin. Overall, these achievements we made in real-time processing, optical localization, system adaptability, and anatomical feature extraction contribute to the application of our real-time volumetric free-hand ultrasound imaging system, particularly for large-sized organs.

## II. METHODS

### A. Data Acquisitions

Our proposed system uses a camera to localize an optical marker tied to the ultrasound probe to acquire the spatial position of the ultrasound probe. The ultrasound machine display screen is transmitted in real-time to a computer display through a frame grabber, and we obtain the corresponding B-mode image data by taking screenshots.

In the first stage, we selected 10 healthy volunteers for 3D ultrasound imaging of the spine to analyze the reliability of the imaging system. In the second stage, we recruited 6 scoliosis patients at the Department of Spine Surgery, Affiliated Drum Tower Hospital, Medical School of Nanjing University, China. The study was approved by the institutional ethics committee. All participants were fully informed about the study procedures and provided written informed consent before participating. Two system developers performed 3D ultrasound imaging of the spine on the participants, who also underwent a whole-spine radiographic examination within two weeks.

### B. Calibration and Positioning of Ultrasound Probes

Since the acquisition of B-mode images requires a more complex process, there is a delay between the screenshot and the ultrasound probe localization data, which makes the two unable to correspond. To address the issue of mismatch between two-dimensional (2D) ultrasound images and spatial localization data, we employed the temporal calibration method proposed by Treece et al. [33]. This method determines the system delay by comparing the vertical motion of a line segment in a 2D image with the vertical motion of the ultrasound probe captured by the localization system.

The spatial localization of the ultrasound probe as well as the calibration work relies on the optical calibration technique proposed previously by our research group [34]. By combining Augmented Reality University of Cordoba (ArUco) code with circular arrays, the optical marker can improve the accuracy of optical localization to 0.02mm. The positional relationship between the optical marker and the ultrasound image is based on a multi-layer N-line calibration technique. Ultimately, the entire system enables spatial localization of 2D ultrasound images with an accuracy of 1 mm.

Furthermore, we optimized the optical localization process to enhance the data acquisition frame rate and minimize sampling time. This was achieved by capturing the rough area of the optical marker in the previous frame obtained from the camera. Subsequently, we performed a cropping operation on the current acquired image, focusing only on the captured rough area. By employing this approach, we only need to recognize one-third or even a smaller area of the original image, thus increasing the rate of localization for the optical marker to 30ms.

### C. Real-time 3D Imaging

After carefully reviewing the findings of Solberg et al. [35] and taking into account factors such as time consumption and imaging effectiveness, we decided to utilize the pixel-nearest neighbor (PNN) algorithm for the reconstruction process. This algorithm was selected based on its proven capabilities and suitability for our needs, ensuring desirable results in terms of both reconstruction speed and geometry measurements of large-sized targets.

To achieve real-time 3D imaging, this study employed an incremental imaging method with algorithmic acceleration using the CUDA Toolkit (CUDA, NVIDIA, USA) and multi-threaded processing. Typically, in cases where the amount of accumulated data is large, it takes more than 3 minutes to reconstruct and visualize all the data at 1mm resolution. The incremental reconstruction method splits the image reconstruction process to reconstruct only the newly acquired data each time, reducing the amount of data in a single reconstruction process. Moreover, in the visualization process, only the areas that need to be updated are processed, which reduces the rendering workload and saves computational time substantially. We use multi-threading processing to carry out the data acquisition, reconstruction, and rendering process simultaneously so that the system has the ability of real-time imaging. During the spine scanning process, the operator held an ultrasound probe and scanned three columns side by side from the cervical to the lumbar spine. Fig. 1 demonstrates the real-time 3D imaging of the spine at the 40s, 80s, and 120s of data acquisition time. Once the data acquisition process was completed, real-time imaging ceased, and the entire image was automatically saved to the computer as slices along the x-axis.

### D. Feature Extraction

When performing spine data acquisition on the human body, we found that there would be some high-brightness tissues above the spine, and the depth of the bone surface varied greatly in different locations of the spine, as shown in Fig. 2. These made it difficult for us to show the anatomy of the spine. To address this problem, we proposed two new tissue segmentation algorithms. One is used for real-time 3D ultrasound imaging, which solves the problem that the spine of the same person has different depths at different locations. The other algorithm is used for the 3D visualization process after data acquisition, which is simpler to use without parameter adjustment.

In the first tissue segmentation algorithm employed in real-time imaging, we utilize the anatomical structure of the human back. By calculating the spatial distance between the ultrasound image and the neck where the initial B-mode ultrasound image is located, we dynamically adjust the image's depth of cut based on the acquired probe localization data. The process of the tissue segmentation algorithm is described in detail as shown in Table I.

The transformation matrix $T_W^{Cam}$ translates a voxel from the world coordinate system $\{W\}$ to the camera coordinate system $\{Cam\}$. The transformation matrix $T_T^W$ translates a voxel from the transducer coordinate system $\{T\}$ to the world coordinate system $\{W\}$. $K$ is a custom constant used to translate spatial distance relationships into image cut depths, and $D$ is another constant that represents the initial value of the image cut depth. ***Spine*** means 3D anatomical data of the spine in the camera coordinate system. This tissue segmentation algorithm takes into account the characteristics of the human back, where the tissue above the central spine is thinner and thicker on both sides. By comparing the probe position information, we can roughly achieve a shallower cut in the center and a deeper cut on the sides. It has the advantage of low computational effort and complexity, making it suitable for real-time imaging systems.

In the other tissue segmentation algorithm for the post-acquisition process, as shown in Fig. 4, we identify the position of the spine in a 2D ultrasound image, use median filtering to exclude error data and reconstruct the spinal anatomy based on the cut image. The process of the tissue segmentation algorithm is detailed in Table II.

TABLE I

**Algorithm 1** Tissue Segmentation Algorithm Used in Real-time Imaging

**Input:** *The screen shots $Img$, the length of back $L$, the transformation matrices $T_W^{Cam}$ and $T_T^W$.*
**Output:** *The visualization of spinal anatomy.*
*Step1. Extract the B-mode images $Img_B$ from $Img$.*
*Step2. Calculate the position of the upper-left pixel of the initial B-mode image in the camera coordinate system $P_{c1}$:*
$$P_{c1} = T_{W1}^{Cam} T_T^W$$
*Step3. Calculate the position of the upper-left pixel of the $n$th B-mode image in the camera coordinate system $P_{cn}$:*
$$P_{cn} = T_{Wn}^{Cam} T_T^W$$
*Step4. Calculate the positional relationship of the $n$th B-mode image relative to the middle of the back along the x-axis (spine direction) and get the corresponding depth of cut $Cut_n$:*
$$Cut_n = K \left| P_{c1}.x - P_{cn}.x - \frac{L}{2} \right| + D$$
*Step5. Cut the image $Img_B$ using the corresponding depth of cut to obtain the image $Img_C$.*
*Step6. Reconstruct image $Img_C$ using $T_W^{Cam}$ and $T_T^W$:*
$$Spine = T_W^{Cam} T_T^W Img_C$$
*Step7. Visualize the spinal anatomy using The Visualization Toolkits (VTK, Kitware, USA).*

TABLE II

**Algorithm 2** Tissue Segmentation Algorithm Used After the Acquisition Process

**Input:** *The screen shots $Img$, the transformation matrices $T_W^{Cam}$ and $T_T^W$.*
**Output:** *The visualization of spinal anatomy.*

*Step1. Extract the B-mode images $Img_B$ from $Img$.*
*Step2. Extract the contours at the bottom of the image $Img_B$ with large gradient variations and high pixel values.*
*Step3. Obtain the required depth of cut for each image by applying median filtering to all extracted contour information.*
*Step4. Cut the image $Img_B$ using the corresponding depth of cut to obtain the image $Img_C$.*
*Step5. Reconstruct image $Img_C$ using $T_W^{Cam}$ and $T_T^W$ :*
$$Spine = T_W^{Cam} T_T^W Img_C$$
*Step6. Visualize the spinal anatomy using the VTK.*

In Step 2 of Table 2, we extracted contours with higher gradient and brightness values from the lower section of the B-mode ultrasound image to identify 2D anatomical location information. This approach is derived from the observation that the bone cortex appears bright in ultrasound imaging, while the bone itself is not penetrable to ultrasound, resulting in lower brightness beneath its surface [14]. By adaptively adjusting the depth of cut for different body types and different spinal positions in the same body, the problem of incomplete spinal anatomical information due to the fixed depth of cut is solved, and the interference of tissues above the bones on the ultrasound image is reduced. Moreover, unlike the fixed-depth method, which requires the probe to be oriented in the same direction during the scanning process, our method allows operators to appropriately tilt the ultrasound probe, which is more in line with the doctors' usage habits and suitable for clinical applications.

## III. EXPERIMENT AND RESULT

### A. System Components

A schematic of the system that can be used to realize real-time volumetric free-hand ultrasound imaging is shown in Fig. 5. The system consists of an ultrasound imaging device (Acuson s3000, Siemens, Germany) with a convex array transducer (6C1, Siemens, Germany). On this transducer, we attach an optical marker (a flat glass board of 80mm*60mm with 4 Augmented Reality University of Cordoba patterns and 44 black circles, the diameter of the circles is 4 mm and the axial and lateral distance between two circles is 6 mm.) by using a self-made immobilization device and perform spatial localization of the transducer using a high-resolution camera (MV-CH650-90XM, Hikrobot, China). A frame grabber (HV-HCA25P, CreHiVi, China) is installed on a high-performance computer with an Intel Xeon E5-2667 v4 processor and an NVIDIA Quadro P6000 graphics card to transmit ultrasound images at 30 frames per second.

In the diagnosis of scoliosis, which is commonly performed on adolescents or even younger children, the four-point immobilization device does not provide effective immobilization during the scanning. This is because this type of immobilization method relies on the individual's subjective ability to move forward and support their entire body on the four points of contact. During our data collection on younger patients using this device, the patient usually had a large displacement after one minute. To solve the problem of the movement of younger children during spinal data acquisition, we designed a new human immobilization device. The device is secured using a porous plate with a hole spacing of 2cm, allowing it to accommodate individuals of different body sizes. Volunteers are immobilized by clamping the device under the armpit. Additionally, the device stabilizes the crotch position in four directions, minimizing movement of the lower part of the body. It has been confirmed through multiple experiments that this design restricts body movement in more directions without altering the shape of the human spine, and is clinically more suitable for younger children.

### B. Experiments and Results

We set the system sampling rate to 30 fps to reach the upper limit of the frame grabber. The center frequency of the convex array probe was set to 4.5MHz, and the imaging depth was set to 7cm to satisfy the need of spine imaging for people with different body mass index (BMI). The specific scanning area extended from the seventh cervical vertebra (C7) to the last lumbar vertebra (L5), including a 50mm width region on each side. The collected B-mode images had a size of 559×699 pixels. The amount of data collected for each patient was proportional to their body size, with a volunteer of 170cm tall resulting in approximately 3000 images.

To assess the severity of scoliosis, the Cobb angle [36] of the spine is usually measured clinically on radiographs. However, the presence of bony structures such as spinous processes above the vertebral plate causes interference, making it difficult to image the vertebral plate. Therefore, it is challenging to locate the position of the vertebral plate in 3D ultrasound images for the measurement of the Cobb angle. In this study, we used UCA instead of the Cobb angle to assess the severity of scoliosis, the feasibility of which was demonstrated [21]. In Fig. 6, the vertebral processes of the thoracic spine are highlighted with green circles. In the lumbar spine region, the lumps formed by the combination of partial shadows from the bilateral inferior articular processes, the vertebral plate, and the superior articular processes of the lower vertebrae are indicated with blue circles [37]. By comparing the images within the red frames, it can be observed that the imaging results obtained from the volume projection imaging (VPI) algorithm fixed at a depth of 35mm fail to accurately distinguish the location of the lumps. On the other hand, the VPI algorithm fixed at a depth of 45mm encounters difficulties in localizing the thoracic vertebrae. We made several attempts between 35-45mm depths, but it was difficult to visualize each lump in the lumbar spinal structures in the images. In contrast, our proposed segmentation algorithm 1 can balance the quality of spine imaging in real-time imaging for both the thoracic and lumbar regions. The proposed segmentation algorithm 2 can reveal the spinal structure of each part more clearly after data acquisition, especially in the part of the yellow circle in the image, which effectively helps to measure the UCA.

We compared the 3D ultrasound imaging results with the corresponding X-ray images which correctly matched each other in large-scale parameters such as total length, and intercostal distance. The curvature of the spine was independently measured by three experienced spine surgeons with more than 15 years of work experience. Fig. 7(a) shows how surgeons usually diagnose scoliosis by the Cobb angle measured in X-ray images. We compared the results of 3D ultrasound imaging with X-ray imaging and measured the UCA in the ultrasound image as shown in Fig. 7(b). The results of UCA in our acquired patient data had a strong correlation with the corresponding Cobb angle as shown in Fig. 8. The mean measurements in terms of Cobb angle and UCA were 39.8°±15.2° and 40.2°±16.8°, with a correlation coefficient of 0.96, which verifies the feasibility of our proposed system. The reliability of the measurements made by the same medical practitioner as well as between different medical practitioners was assessed using intraclass correlation coefficients (ICC). Data from 6 scoliosis patients were analyzed to obtain intra-observer (ICC=0.94) and inter-observer (ICC=0.92) reliability.

## IV. DISCUSSION

This study implements a real-time volumetric free-hand ultrasound imaging system for large-sized organs. The system is built on existing ultrasound imaging equipment used in clinical and can be easily deployed. Experiments on versatile ultrasound devices such as LOGIQ™ E20 (GE, USA), ACUSON S3000™ (Siemens, Germany), and Clover60 Vet (Wisonic, China) demonstrated the suitability of our system for different ultrasound devices. In addition, our proposed system utilizes a free-hand scanning method and is adapted to various types of ultrasound probes commonly found in hospitals, which makes it more flexible than designs using robotic arms and customized ultrasound probes, and therefore more suitable for imaging large-sized organs with complex structures.

In this study, we achieved a spatial localization accuracy within 1mm by using the optical calibration and tracking technique proposed previously by our research group [34]. We also proposed a recognition logic of the optical markers to achieve a 3D ultrasound imaging speed at 30fps to meet the requirement of scanning the whole spine of a volunteer in 2 minutes by employing incremental reconstruction and rendering techniques, parallel computing techniques to accelerate the data processing. These improvements allow our system to image targets quickly and accurately, laying the groundwork for future practical applications in hospitals and other scenarios.

Furthermore, this study introduced optimizations for practical application scenarios such as scoliosis. Firstly, to address the issue of body movement during prolonged scanning, we not only shortened the data acquisition time but also designed a new human immobilization device to minimize the movement of the human body during the data acquisition process. Additionally, we proposed a new solution for extracting spinal anatomical features. For real-time imaging, we use a segmented linear cutting approach that adjusts the cutting depth based on the location of the image in the spine. This cutting method better conforms to the human anatomy compared to the fixed cutting depth of the VPI algorithm, allowing for clearer visualization of anatomical features at different positions along the spine. However, this cutting method limited adaptability to individuals with different body types and BMIs, requiring parameter adjustments during practical use, which could be challenging for medical practitioners who are users of the system. Therefore, we propose another method for extracting spinal anatomical features, which visualizes the 3D anatomical information of the spine after the data acquisition process. This algorithm dynamically adjusts the cutting depth based on the approximate location of the bones in the acquired images, which eliminates the need for system users to adjust the parameters of the algorithm according to different populations, reducing the difficulty of using the system and making it more suitable for clinical applications.

We also attempted real-time 3D ultrasound imaging of other large-sized targets. As shown in Fig. 9, we imaged the calf and segmented the musculus gastrocnemius from it. This type of image is expected to assist doctors in diagnosing muscular dystrophy or other related diseases."

## V. CONCLUSION

In this study, we propose a real-time volumetric free-hand ultrasound imaging system, which can be built based on multiple types of ultrasound devices with a simple assembly process, and is expected to be applied to clinical scenarios where equipment needs to be brought out for out-of-door medical checkups. The system uses the optical localization method previously proposed by our research group, which can be used to spatially localize the ultrasound probes with high accuracy in hospitals and other environments with strong electromagnetic interference. Moreover, we improved the recognition process of optical markers, together with parallel computing techniques, to achieve data acquisition and real-time visualization, which not only helps medical practitioners diagnose the condition promptly but also provides them with real-time 3D ultrasound images to guide them during the surgery. Successful 3D ultrasound imaging of the spine in different populations confirms the feasibility of the system for real-time 3D ultrasound imaging of large-sized organs. We hope to further apply it to the ultrasound imaging of other large-sized organs in clinical.

APPENDIX

Video 1 shows the real-time 3D ultrasound imaging of a volunteer's spine at The Affiliated Hospital of Nanjing University Medical School.